# Status of the CP-PACS Project

A. Ukawa[a]

[a]Institute of Physics, University of Tsukuba, Tsukuba, Ibaraki 305, Japan

The CP-PACS Project, which started in April 1992, is a five-year plan to develop a massively parallel computer for carrying out research in computational physics with primary emphasis on lattice QCD. This article describes the architectural design of the CP-PACS computer, the entire computing system including the front end and mass storage, and results of benchmarks for the expected performance for lattice QCD applications.

## 1. THE CP-PACS PROJECT[1]

The project name CP-PACS stands for *Computational Physics–Parallel Array Computer System*. The name was chosen to reflect the two phases of the project – development of a massively parallel computer optimized for physics problems describable in terms of space-time fields, and subsequent research with it in several key areas of computational physics with primary emphasis on lattice QCD.

The preparation for the project was initiated in the early summer of 1990. The proposal, submitted to the Ministry of Education, Science and Culture of the Japanese Government, was approved in March 1991. The project formally started in April 1992 with the funding of 1.5 billion Yen, to be spread over a five-year period between fiscal 1992 and 1997, allocated from a special category of the Grant-in-Aid of the Ministry of Education, Science and Culture for promoting innovative and fundamental research in science. Concomitant with the start of the project, Center for Computational Physics was founded at University of Tsukuba in April 1992 in order to serve as a base for a collaborative effort between physicists and computer scientists for the development of the CP-PACS computer and its utilization for research in computational physics.

The project currently consists of 28 members as listed in Table 1. This represents a significant expansion over the predecessors of the project, the PAX Project(1977-1986) of T. Hoshino[2] and the QCDPAX Project(1987-1989) of Y. Iwasaki and T. Hoshino[3], which, carried out by a handful of people, pioneered the development and application of parallel computers to scientific calculations in Japan. As Table 1 clearly shows, the CP-PACS Project is a multi-disciplinary effort toward the advancement of computational physics encompassing not only several branches of physics but also computer science to develop parallel computers best suited for such applications. The project is headed by Y. Iwasaki and the development of the computer system is led by K. Nakazawa.

Development of a massively parallel computer requires advanced semiconductor technology. We discussed the aim of the project with a number of manufacturers and invited proposals in the period of 1991–1992. We have selected Hitachi Ltd. for the manufacturing of the CP-PACS computer through a formal bidding process in the early summer of 1992. The company has since been working with us in a close collaboration both on the hardware and software development of the CP-PACS computer.

## 2. CP-PACS COMPUTER
### 2.1. Design parameters

The target specification of the CP-PACS computer is summarized in Table 2. The basic strategy we have adopted for the design is the usage of a fast RISC micro-processor for high arithmetic performance at each node and a linking of nodes with a flexible network so as to be able to handle a wide variety of problems in computational physics. The unique features of the CP-PACS computer reflecting these goals are represented by the special node processor architecture called *pseudo vector processor based on slide-windowed*

2Table 1
CP-PACS Project members

| hardware | software | particle physics | astrophysics | solid state physics |
| --- | --- | --- | --- | --- |
| K. Nakazawa[a] | I. Nakata[a] | Y. Iwasaki[g] | M. Miyama[k] | S. Miyashita[l] |
| H. Nakamura[a] | Y. Yamashita[a] | A. Ukawa[g] | T. Nakamura[i] | M. Imada[m] |
| T. Boku[a] | Y. Oyanagi[c] | K. Kanaya[h] | M. Umemura[h] | |
| T. Hoshino[b] | T. Kawai[d] | S. Aoki[g] | Y. Nakamoto[g] | |
| T. Shirakawa[b] | M. Mori[e] | T. Yoshie[h] | | |
| K. Wada[a] | Y. Watase[f] | M. Fukugita[i] | | |
| | S. Ichii[f] | M. Okawa[j] | | |
| | | N. Ishizuka[g] | | |
| | | H. Kawai[j] | | |

[a] Institute of Information Sciences and Electronics, University of Tsukuba
[b] Institute of Engineering Mechanics, University of Tsukuba
[c] Department of Information Science, University of Tokyo
[d] Department of Physics, Keio University
[e] Department of Engineering, University of Tokyo
[f] Data Handling Division, KEK
[g] Institute of Physics, University of Tsukuba
[h] Center for Computational Physics, University of Tsukuba
[i] Yukawa Institute for Theoretical Physics, Kyoto University
[j] Theory Division, KEK
[k] National Observatory
[l] Institute of Humanities and Environments, Kyoto University
[m] Institute of Solid State Physics, University of Tokyo

registers ($PVP\text{-}SW$)[4] and the choice of a three-dimensional crossbar network.

The computer is scalable up to 2048 nodes. In order to exploit this potential, a request is being made for an additional funding for a grade-up of the CP-PACS computer to the Ministry of Education, Science and Culture .

## 2.2. Node processor
### 2.2.1. PVP-SW[4]

A wide class of numerical calculations in scientific problems is well suited for vector processing. Vector enhancement of RISC micro-processors, however, faces a number of engineering problems. One of the problems is that of a large memory latency, namely the time needed to fetch data from main memory to registers is much larger than the processor machine cycle time. The bandwidth between memory and processor also has to be sufficiently high so that the processor is kept supplied with new data at every machine cycle. Finally the number of floating point registers should be large in order to handle complex calculations in scientific applications.

It is worth noting that a cache memory widely used in commercial RISC micro-processors is inadequate to overcome these problems for large-scale scientific applications since data size easily exceeds the cache memory capacity and that data generally have a low temporal locality.

The slide-windowed registers with preload features[4] resolve the problem of large memory latency and the necessity of a large number of floating point registers while keeping upward compatibility with commercial RISC micro-processors, in which usual arithmetic instructions can specify only a limited number of registers such as 32.

A schematic illustration of slide-windowed registers is given in Fig. 1. The number of physical floating point registers is some large number $m > 32$, while the number of logical registers specifiable by usual instructions is 32. The logical registers are divided into $g$ *global* registers and $32-g$ *local* registers. The global registers are fixed at the top of the physical registers. The local registers, on the other hand, can slide along the rest of the physical registers with the *slide length* taking any value specifiable in user application

Table 2
Target specification of the CP-PACS computer

| | |
|---|---|
| peak speed | 300Gflops(64 bit data) |
| main memory | 64GB |
| parallel architecture | MIMD with distributed memory |
| number of nodes | 1024 |
| node processor | |
|   architecture | HP PA-RISC1.1+PVP-SW |
|   #FP registers | 128 |
|   clock cycle | 150MHz |
|   1st level cache | 16KB(I)+16KB(D) |
|   2nd level cache | 512KB |
| network | 3-d crossbar |
|   node array | $8 \times 17 \times 8^*$ |
|   through-put | >150MB/sec |
|   latency | <7$\mu$sec |
| distributed disks | 3.5" RAID-5 disk |
|   total capacity | 538GB |
| software | |
|   OS | UNIX micro kernel |
|   language | FORTRAN, C, assembler |

*including nodes for disk I/O

programs. The totality of the local registers at a particular position and the global registers is called a *window*.

Suppose that the logical registers are at the window $i$. While carrying out computations using the registers in this window, the processor can issue *preload* instructions which fetch data from memory to registers in any forward window $j \geq i+1$. The processor can also issue *poststore* instructions which store data in any previous window $k \leq i-1$ to memory. Instructions that follow a *preload* or a *poststore* do not wait completion of the *preload* and *poststore* instructions. For $m$ sufficiently large, one can clearly arrange the value of $j$ so that the data already reside in registers when the window is shifted to $j$, thereby effectively reducing the memory latency to null.

A high bandwidth between memory and registers is an independent requirement. A conventional method for solving this problem is to pipeline the memory with multiple interleaved banks, which we adopt for our processor.

Let us illustrate how pseudo vector processing works with the slide-windowed registers by a simple example of the DAXPY loop $z_i = a \times x_i + y_i, i = 1, \cdots, n$. For each loop cycle, we need to load $x_i$ and $y_i$, carry out a multiplication $(a, x_i) \to a \times x_i$ and an addition $(a \times x_i, y_i) \to a \times x_i + y_i$, and finally store $z_i$.

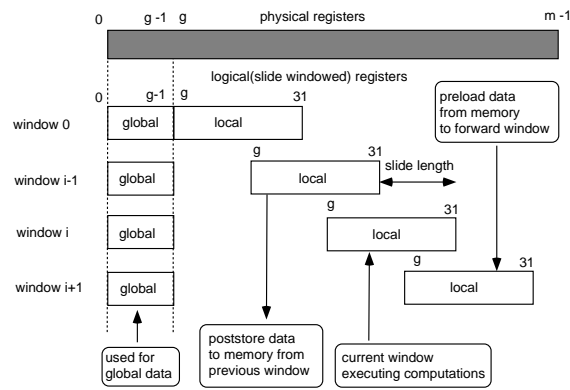

Figure 1. Structure of slide-windowed registers.

Figure 2. A possible instruction scheduling for DAXPY loop with PVP-SW.

The idea to vectorize these operations is to interleave instructions from neighboring loop cycles so that the processor machine cycles are filled up with instructions. This can be carried out by an application of the software pipelining technique called module scheduling[5,4], which results in the scheduling shown in Fig. 2. Here we assumed a memory latency of 7 machine cycles and a latency of 3 machine cycles for the completion of an addition or a multiplication. Branch instructions are omitted for simplicity as well as the superscalar features.

We observe that there are two *preloads*, one *poststore*, one multiplication and one addition for each window as required. However, the instructions in a window generally refer to different loop cycles. The memory latency is completely hidden by the use of preloading to the next window. Compared to 15 machine cycles needed for a scalar code, a loop cycle is reduced to $5+1$ machine cycles with the last machine cycle necessary for the *window switch* instruction.

To write an executable assembler code we still need to decide the *slide length* and allocate registers. The minimum value of *slide length* equals the number of *preloads*, which equals 2 for the present case. The constant $a$ common to all the loop cycles is stored in a global register. Allocation of $x_i$ and $y_i$ to local registers can be worked out by a coloring technique[6]; we refer to Ref. [4] for details .

### 2.2.2. Implementation

We employ the Hewlett-Packard PA-RISC 1.1 architecture for the base architecture of the node processor. The instruction set is enhanced to include *preload*, *poststore* and *window switch* for incorporating the PVP-SW. The peak speed of the processor is 300Mflops for 64 bit data with the machine cycle of 150MHz.

A detailed examination of the performance of the enhanced architecture for the Livermore kernels[4] has shown that a memory latency of up to 70 machine cycles can be hidden if the number of physical registers $m$ exceeds 96. Benchmarks for the core of the minimal residual solver for the Wilson quark action yielded a somewhat larger estimate of $m \approx 110$ (see Sec. 4). We have chosen $m = 128$ for the number of slide-windowed physical registers.

The processor has 64Mbyte of DRAM for main memory, which is pipelined with multiple interleaved memory banks. The bandwidth is designed to be roughly one 8byte word for each machine cycle.

The processor also has two levels of cache memory; the on-chip first level cache for data and instructions separately, each with a capacity of 16Kbyte, and an off-chip second level cache of at least 512Kbyte.

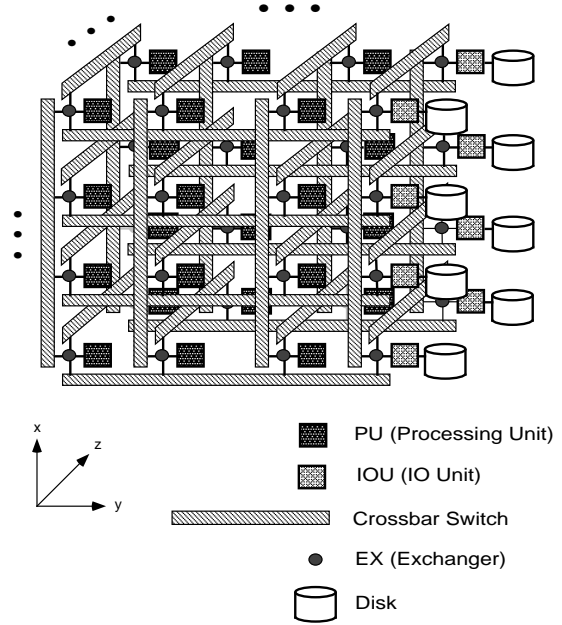

Figure 3. Schematic diagram of CP-PACS.

### 2.3. Network

The network of the CP-PACS computer is best illustrated by a schematic diagram shown in Fig. 3. The 1024 processors are arranged in a three-dimensional $8 \times 16 \times 8$ array. To connect the processors together, a number of crossbar switches (shaded bars) are placed in $x, y$ and $z$ directions. The crossbars for different directions are connected at each crossing point by a router, called an exchanger (shaded circles), which is a $4 \times 4$ crossbar itself. A processing unit (PU) consisting of a processor with its memory(shaded squares) is attached to each exchanger.

Inter-node communication is made by message passing. Transfer of data within the network proceeds via wormhole routing through the exchangers. The direction of routing is fixed to $x \to y \to z$ to avoid deadlocks. The network also supports broadcast and barrier synchronization.

This represents a very flexible architecture allowing data to be sent from a processor to any other processor in at most three steps. The network also allows a hardware barrier to be set up across several two-dimensional planes so that the processor array can be divided into a set of independent subarrays. For lattice QCD applications, a crucial advantage of the architecture is that virtually any lattice size can be simulated, subdividing the processor array if necessary. Hence finite lattice size effects, which have proven very important in recent lattice QCD studies, could be examined in detail.

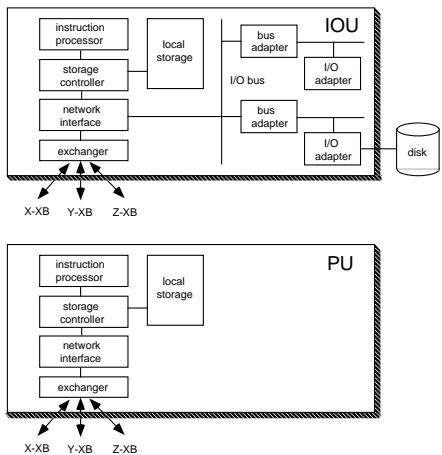

Figure 4. Schematic diagram of IOU and PU.

Transfer of data is controlled by a UNIX micro kernel which resides in each processor. There is a fast mode of communication in which data in the user application area of memory are directly sent to the network interface without buffering to the system memory area, and are directly written into the memory of the receiving processor also without system buffering. The fast mode supports a block-strided transfer, i.e., a set of data regularly spaced in memory addresses can be sent by a single command.

The bandwidth of each crossbar is at least 150Mbyte/sec. The latency, namely the setup time for sending and receiving data, is less than $7\mu sec$ for the fast mode combining those due to hardware and software.

### 2.4. Distributed disks

The CP-PACS computer is designed to have over 500Gbyte of disk space for storing intermediate results. Important practical problems with such a large capacity is the reduction of I/O time between processor memory and disks, and the guard against loss of data due to disk failures.

These problems are solved by distributing disks across the $x-z$ plane of the processor array and employing the RAID-5 technology. As shown in Fig. 3, an I/O unit (IOU) is attached at the end of each crossbar in the $y$ direction, to which is connected a 3.5 inch RAID-5 disk with a capacity of 8.4Gbyte through a SCSI-II bus. Thus the CP-PACS computer has $8 \times 8 = 64$ IOU's with a total peak throughput of $64 \times 10 = 640$Mbyte/sec and a total disk space of 538Gbyte. To accommodate the I/O units, the crossbars in the $y$ direction are enlarged to $17 \times 17$ crossbars. A schematic diagram of IOU is given in comparison to that of PU in Fig. 4.

### 2.5. Software development
#### 2.5.1. Operating system

The CP-PACS computer is run on the UNIX operating system. Each node processor, however, carries only a micro kernel of UNIX in order to save memory for user application programs and to avoid a degradation of performance. The kernel handles memory control, inter-node communications, process scheduling, interruptions and I/O. Full UNIX operating system is implemented on one of the IOU, named SIOU, which controls parallel processes through the network.

The operating system has several new functions added for parallel processing; software partitioning of the processor array so that independent programs may be run on different partitions, and generation of processes over a user-specified number of nodes to execute a parallel program.

The file system is logically structured to form a single tree for the entire CP-PACS computer. A logical single file, however, may be physically divided across the disks so as to reduce I/O overheads. These divisions and allocations are made automatically by the operating system.

#### 2.5.2. Programming environment

The programming language of the CP-PACS computer is FORTRAN, C and assembly language. An assembler code can be included as a subroutine in a FORTRAN or C code in order to maximize the performance. FORTRAN and C compiler are being developed incorporating the PVP-SW enhancement. The technique of modulo scheduling and register coloring can be employed for the compiler algorithm (see Sec. 2.2.1), and a prototype of a FORTRAN complier has already been written.

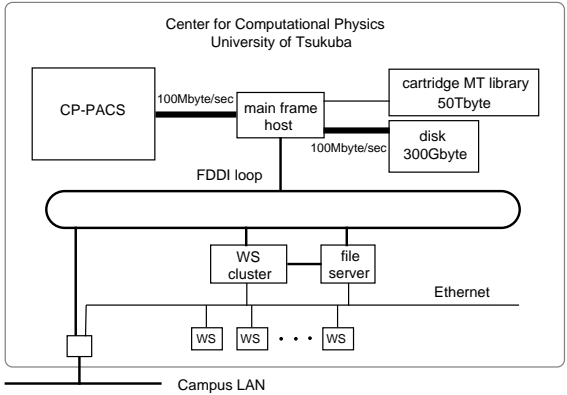

Figure 5. Plan for the computing system at the Center for Computational Physics.

## 3. FRONT END AND MASS STORAGE

We show in Fig. 5 the plan for the computing system at the Center for Computational Physics. In order to store data produced by the CP-PACS computer, two levels of mass storage are considered: 300Gbyte of disks and up to 50Tbyte of cartridge magnetic tape library. Transfer of data is mediated by a main-frame host computer using a high speed link with the peak throughput of 100Mbyte/sec. The host also serves as a front end for controlling the CP-PACS computer and its job scheduling. Program development and data analyses are carried out on the host and work stations connected by an FDDI and Ethernet local area network, which also serves as a connection to the outside of the Center through the University of Tsukuba Campus LAN.

## 4. LATTICE QCD BENCHMARKS
### 4.1. Wilson minimal residual solver

Let us describe the expected performance of the CP-PACS computer for lattice QCD calculations based on benchmarks assuming 2048 processors.

The core of the minimal residual solver for the Wilson quark action may be symbolically written as $H = G - \kappa * U * G$ where $G$ is the solution vector, $U$ the link matrices and $\kappa$ the hopping parameter. For each Dirac index, the calculation of $V = U * G$ requires 30 loads, 6 stores, 36 additions and 36 multiplications. Ideally these operations can be executed in 36 machine cycles with the superscalar architecture of PA-RISC 1.1. The calculation of $G - \kappa * V$, on the other hand, is less balanced requiring 56 machine cycles to accommodate 40 loads, 8 stores, 56 additions and 16 multiplications. Thus the theoretical maximum of the efficiency defined by $\#operations/(2 \times \#machine\ cycle)$ is $(36 + 36 + 56 + 16)/(2 \times (36 + 56)) = 82\%$.

In collaboration with Hitachi Ltd., we have developed a hand-optimized assembler code for this part of the solver including the PVP-SW features. The code uses a slide length of 10 and preloading to a window 6 times forward to the current window. The efficiency turned out to be 75% for an infinite lattice and 74% for a $64^4$ lattice divided into $8 \times 4 \times 4 \times 64$ sublattices on an $8 \times 16 \times 16$ processor array. The degradation of efficiency stems from window switch instructions and address calculations.

The efficiency above does not take into account the effect of memory bank conflict. This effect was estimated by a simulation program which incorporates the actual structure of memory. It was found that the loss of efficiency is less than 1% if the number of physical registers is 110 or more, which is satisfied with 128 registers adopted for the processor.

The communication overhead was estimated assuming a bandwidth of 150Mbyte/sec and a latency of $7\mu$sec for the crossbar network. The efficiency decreases from 74% to 57% for a $64^4$ lattice. The loss primarily comes from the bandwidth, which we hope to improve for the actual machine.

A similar estimate for the Kogut-Susskind quark action yielded an efficiency of 81% for a single processor and 67% including communications for a $64^4$ lattice.

For the Wilson minimal residual solver with a red/black preconditioning we estimate the efficiency including communications to be 63%, of which 82% are for floating point operations and 18% for communications. The actual execution time for a single iteration of the solver is expected to take 0.075sec for a $64^4$ lattice.

### 4.2. CPU time estimates for physics runs

Representative estimates of CPU time for lattice QCD runs based on the benchmarks of

Table 3
Estimate of CPU time for physics runs on the CP-PACS computer assuming 2048 processors.

(a) quenched Wilson spectroscopy

| assumptions: | |
|---|---|
| size | $64^4$ |
| $m_\pi/m_\rho$ | 0.9 – 0.4 in steps of 0.1 |
| # conf. | 500 |
| CPU time (days) | |
| gauge sweep | 22 |
| quark propagator and observables | 36 |
| total | 58 |

(b) full QCD configuration generation

assumptions:
HMC(Wilson) or hybrid R(KS)
red/black MR(Wilson) or CG(KS)
# trajectories = 2000

| CPU time (days) | | | | | |
|---|---|---|---|---|---|
| size | action | \multicolumn{4}{c}{$m_\pi/m_\rho$} |
| | | 0.6 | 0.5 | 0.4 | 0.3 |
| $32^3 \times 64$ | W | 65 | 109 | 187 | |
| | KS | 6 | 12 | 34 | |
| $48^3 \times 12$ | W | | 64 | 111 | 210 |
| | KS | | 7 | 22 | 63 |

Sec. 4.1 are listed in Table 3. For quenched QCD we expect to carry out exhaustive measurements with the quark mass ranging from heavy to light for a lattice size of typically $64^4$. Simulations carried out for three values of lattice spacing to control the continuum limit should take about half a year of CPU time.

Full QCD simulations are much more time consuming. The situation is particularly severe for the Wilson quark action. Studies are currently conducted to search for an efficient parallelization of preconditioning methods such as the incomplete LU decomposition which is known to reduce the CPU time by a factor of five or more for vector computers.

## 5. SUMMARY

In this article we have described the CP-PACS Project and the architecture of the planned parallel computer. At present the hardware design is near completion and checks by simulation are well underway. The physical packaging design is also being finalized. A prototype, mainly for the purpose of checking the crossbar network, has been built, and tests of software have been started. Manufacturing of components will start early next year, and assembling of the 1024 node CP-PACS computer is scheduled for completion by the spring of 1996. The start of physics computations is expected by the spring of 1997.

## ACKNOWLEDGEMENTS

I would like to thank the members of the CP-PACS Project, especially Y. Iwasaki and K. Nakazawa, for valuable discussions and for useful suggestions on the manuscript. This work is supported in part by the Grand-in-Aid of the Ministry of Education, Science and Culture (No. 06NP0601, No. 06640372)